\documentclass[aps,prl,twocolumn,superscriptaddress,a4paper,floatfix]{revtex4}
\usepackage[latin1]{inputenc}
\usepackage{graphicx}
\usepackage{amsmath,amssymb}
\usepackage{hyperref}
\usepackage{datetime}
\usepackage{siunitx}
\usepackage{braket}
\usepackage{cleveref}
\usepackage{gensymb}

\begin{document}

\title{Universal pair-polaritons in a strongly interacting Fermi gas}

\author{Hideki Konishi*}
\author{Kevin Roux*}
\author{Victor Helson}
\author{Jean-Philippe Brantut}
\affiliation{Institute of Physics, EPFL, 1015 Lausanne, Switzerland}
\date{\pdfdate}

\begin{abstract}
Cavity quantum electrodynamics (QED) manipulates the coupling of light with matter, and allows for several emitters to couple coherently with one light mode \cite{Haroche:2006aa}. However, even in a many-body system, the light-matter coupling mechanism was so far restricted to one body processes. Leveraging cavity QED for the quantum simulation of complex, many-body systems has thus far relied on multi-photon processes, scaling down the light-matter interaction to the low energy and slow time scales of the many-body problem \cite{Klinder:2015ab,Landig:2016aa,Vaidya:2018aa,Norcia:2018aa}.
Here we report on cavity QED experiments using molecular transitions in a strongly interacting Fermi gas, directly coupling cavity photons to pairs of atoms. The interplay of strong light--matter and strong inter-particle interactions leads to well resolved pair-polaritons, hybrid excitations coherently mixing photons, atom pairs and molecules. The dependence of the pair-polariton spectrum on interatomic interactions is universal, independent of the transition used, demonstrating a direct mapping between pair correlations in the ground state and the optical spectrum. This represents a magnification of many-body effects by two orders of magnitude in energy. In the dispersive regime, it enables fast, minimally destructive measurements of pair correlations, and opens the way towards their measurements at the quantum limit and their coherent manipulation using dynamical, quantized optical fields.
\end{abstract}
\maketitle

\begin{figure}[tb]
	\includegraphics{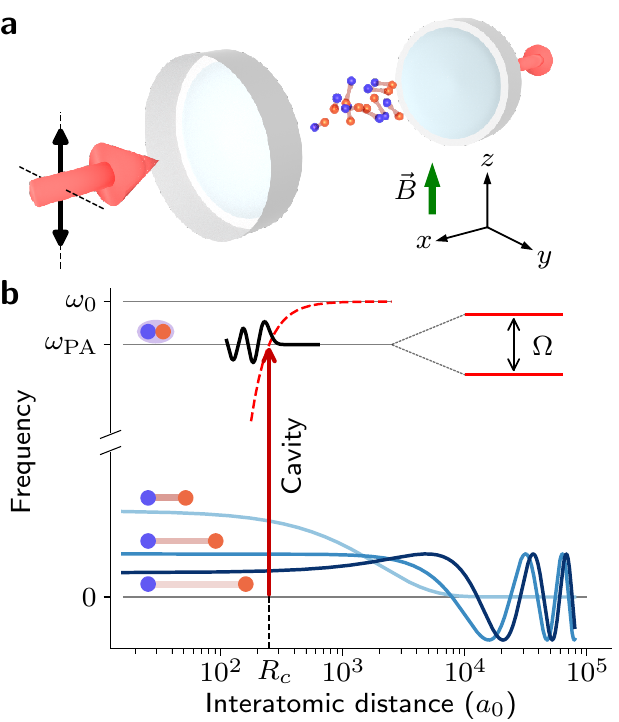}
	\caption{\textbf{Concept of the experiment.} \textbf{a}~A degenerate, strongly interacting two-component Fermi gas is placed inside a cavity. A beam linearly polarized along the bias magnetic field $\vec{B}$ measures the transmission through the cavity. \textbf{b}~Sketch of the two-atoms ground state wavefunction for the BEC, unitary and BCS situations (light blue, blue and dark blue solid lines, respectively) in the ground state, and wavefunction of a molecular state (solid black line) at the frequency $\omega_\mathrm{PA}$ bound in a molecular potential (red dashed line) asymptotically reaching the single atom transition frequency $\omega_0$. The cavity photons at frequency $\omega_\mathrm{PA}$ induce transitions between free atoms and molecular states at the Condon point $R_c$. The collective Rabi frequency $\Omega$ of the process exceeds the rate of dissipation yielding a pair of resolved dressed states in the spectrum.}
	\label{Fig1}
\end{figure}

One of the most striking success of quantum science is the ability to engineer the interaction between light and matter, culminating with the strong coupling regime in cavity quantum electrodynamics (QED)~\cite{Haroche:2006aa,Tanji-Suzuki:2011ac}. Cavity QED is already a corner stone of quantum networks~\cite{Reiserer:2015aa} and quantum information processing~\cite{Krantz:2019aa}. It is now emerging as a new tool for quantum simulation with quantum gases, where it provides unique features such as long-range, collective interactions~\cite{Munstermann:2000aa,Mottl:2012ab,Ritsch:2013aa,Vaidya:2018aa,Norcia:2018aa}, controlled dissipation enabling novel non-equilibrium dynamics~\cite{Dogra:2019aa} and real-time readout~\cite{Mekhov:2007aa}. However, the light--matter interaction was so far limited to the dipole coupling with single atoms, which dominates the energy scales of quantum gases by several orders of magnitudes. As a result, an interplay between many-body physics in a quantum gas and light--matter interactions could only be observed in the dispersive regime, effectively scaling down the light--matter interaction to match that of motional degrees of freedom of atoms.

\begin{figure*}[ht]
	\includegraphics{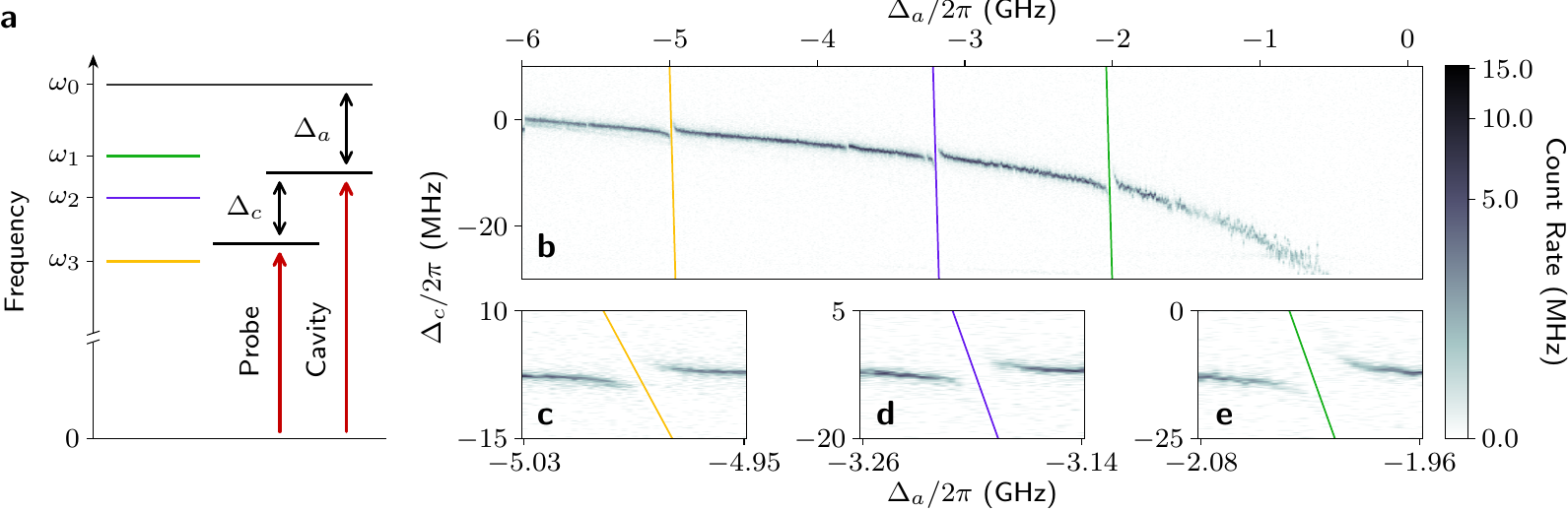}
	\caption{\textbf{Strong coupling on PA transitions.} \textbf{a}~Frequency diagram relevant for the experiment. We address three PA transitions at $\omega_1$, $\omega_2$ and $\omega_3$ below the single atom D2 $\pi$-transition ($\ket{2S_{1/2},m_J=-1/2}\rightarrow\ket{2P_{3/2},m_J=-1/2}$) at $\omega_0$. The cavity--atom and probe--cavity detunings $\Delta_a$ and $\Delta_c$ are independently controlled by tuning the cavity resonance and the probe frequency. \textbf{b}~Cavity transmission spectrum of a unitary Fermi gas below the D2 $\pi$-transition. The solid lines indicate three PA transitions labeled as 1, 2 and 3 as in panel a. A few narrower PA signals can also be seen, which we attribute to PA into other interaction potentials. \textbf{c--e}~Close up view in the vicinity of the three PA lines showing avoided crossing patterns.}
	\label{Fig2}
\end{figure*}

In a many-body system in free space, it is well known that photons can not only be absorbed and reemitted by individual atoms, but also exchanged between atoms yielding a dipole-dipole interaction. Excited molecular states then form in the attractive branches of this interaction potential, yielding photo-association (PA) resonances in the optical spectrum~\cite{Jones:2006aa}. When driving such a PA line, photons couple directly to pairs of free atoms separated by a distance $R_c$, the Condon length of the target molecular state. In quantum gases, PA is weak because $R_c$ is much shorter than the mean interparticle spacing, such that its investigation in strongly interacting systems has been limited to incoherent processes~\cite{Partridge:2005aa,Kinoshita:2005ab,Semczuk:2014aa,Liu:2019aa,Paintner:2019aa}. Coherent Rabi oscillations on ultra-narrow PA transitions have been observed in weakly interacting gases of two-electron atoms~\cite{Yan:2013aa,Taie:2016aa}.

In our experiment, we address PA resonances in a strongly correlated two-component Fermi gas with photons in a high-finesse cavity, as depicted in figure~\ref{Fig1}a~\cite{Roux:2020aa}. Our experiment brings PA in the strong coupling regime, where the interaction between photons and atom pairs overcomes all dissipative processes. In such a regime, pairs, molecules and cavity photons coherently hybridize into composite quasiparticles, pair-polaritons. These coherent excitations inherit from their photonic part the fast dynamics of the resonant light--matter interaction, much faster than that of the many-body physics in the gas, and a weak dissipation channel into the environment enabling direct optical detection. From their matter part, they inherit the universal properties of the short-distance pair correlations of the quantum gas, which thus get imprinted onto the optical spectrum. 

This combination of light--matter and atom--atom interaction is embodied in the collective Rabi frequency $\Omega$ verifying
\begin{multline}
\Omega^2 = \Omega_0^2 \int  d{\bf R} |g({\bf R})|^2 \int d{\bf r} d{\bf r}' f({\bf r})  f^*({\bf r}') \\ 
\left\langle \hat{\psi}_1^\dagger({\bf R} + \frac{{\bf r}'}{2}) \hat{\psi}_2^\dagger({\bf R} - \frac{{\bf r}'}{2})   \hat{\psi}_2({\bf R} - \frac{{\bf r}}{2}) \hat{\psi}_1({\bf R} + \frac{{\bf r}}{2}) \right \rangle,
\label{omega_main}
\end{multline}
where $f(r)$ is an orbital describing the relative motion of the two atoms in the target molecular state, $\Omega_0$ is the single-photon--single-pair Rabi frequency and  $g({\bf R})$ the cavity mode function. $\hat{\psi}_1({\bf r}), \hat{\psi}_2({\bf r})$ are field operators annihilating atoms in states $1$ and $2$ at point $\bf r$. In the linear response regime, the expectation value is taken in the unperturbed thermal state of the gas, and the integral counts the total number of pairs of atoms overlapping with the excited molecular state. Thus, the pair correlations are directly scaled up by $\Omega_0$ such that the low energy physics of the many-body system, which happens at the scale of the Fermi energy, is translated into a measurable difference in the optical spectrum at energy scales two orders of magnitudes larger. Conversely, in weakly interacting systems, the absence of pair correlations prevented so far the observation of PA in the strong coupling regime \cite{Brennecke:2007aa,Colombe:2007aa}.

We use Fermi gases of $^6$Li comprising $4.6\times10^5$ atoms at a temperature $T=0.07\,T_F$ with $T_F$ the Fermi temperature, equally populating the two lowest hyperfine states, denoted as $\ket{1}$ and $\ket{2}$.
A homogeneous magnetic field $B$ set in the vicinity of the broad Feshbach resonance at $832\,$G brings the gas in the strongly interacting regime, where it explores the BEC--BCS crossover~\cite{Giorgini:2008aa,Zwerger:2012aa}. In the PA process, the photons address the two-body wavefunction of atom pairs as sketched in figure~\ref{Fig1}b.
We probe the optical excitation spectrum of the system using transmission spectroscopy, with light linearly polarized along the magnetic field direction (see Methods). The frequency of the probe beam is swept over $40\,$\si{\mega\hertz} in $500\,$\si{\micro\second}, such that light is present in the cavity for about $12.5$ \si{\micro\second}, short compared with the dynamical time scales of the gas.

Figure~\ref{Fig2} shows a typical transmission spectrum for a gas at unitarity, where the detuning $\Delta_a/2\pi$ between the cavity and atomic transition $\ket{2S_\mathrm{1/2},m_J=-1/2} \rightarrow \ket{2P_{3/2},m_J=-1/2}$  spans $6$\,\si{\giga\hertz}. The spectrum is dominated by the strong coupling of photons with the atomic transition at $\Delta_a=0$, yielding the smooth background variation of the transmission resonance with respect to the empty cavity at $\Delta_c=0$, with $\Delta_c$ the detuning of the probe with respect to the empty cavity resonance. Additionally, we observe three successive avoided crossings corresponding to three PA transitions (figure~\ref{Fig2}e, d and c) denoted 1, 2 and 3. The presence of clear avoided crossings confirms that the strong coupling is reached for each of these PA transitions. A simple model (see Methods) suggests that they correspond to vibrational levels $-12,-13$ and $-14$ below the continuum, in the molecular potential resulting from the exchange of $\pi$-polarized photons between atoms. We observed similar patterns for several PA resonances corresponding to bound states in the $1 ^3 \Sigma_g^+$ potential \cite{Abraham:1995aa} for larger $\Delta_a$. 

We fit the spectrum in the vicinity of each PA transition using an analytical model describing a molecular resonance coupled to one cavity mode (see Methods). We extract the value of the Rabi frequency, the location of the molecular states and the dispersive shift due to the coupling to the atomic transitions. The latter is used to directly infer the total number of atoms $N$ on each measurement. The Rabi frequencies extracted for PA resonances 1, 2 and 3 are respectively $2\pi \times 20.7(14)$, $15.6(6)$ and $11.3(2)$\,\si{\mega\hertz}, exceeding both the atomic and the cavity decay rates.

\begin{figure}[t]
	\includegraphics{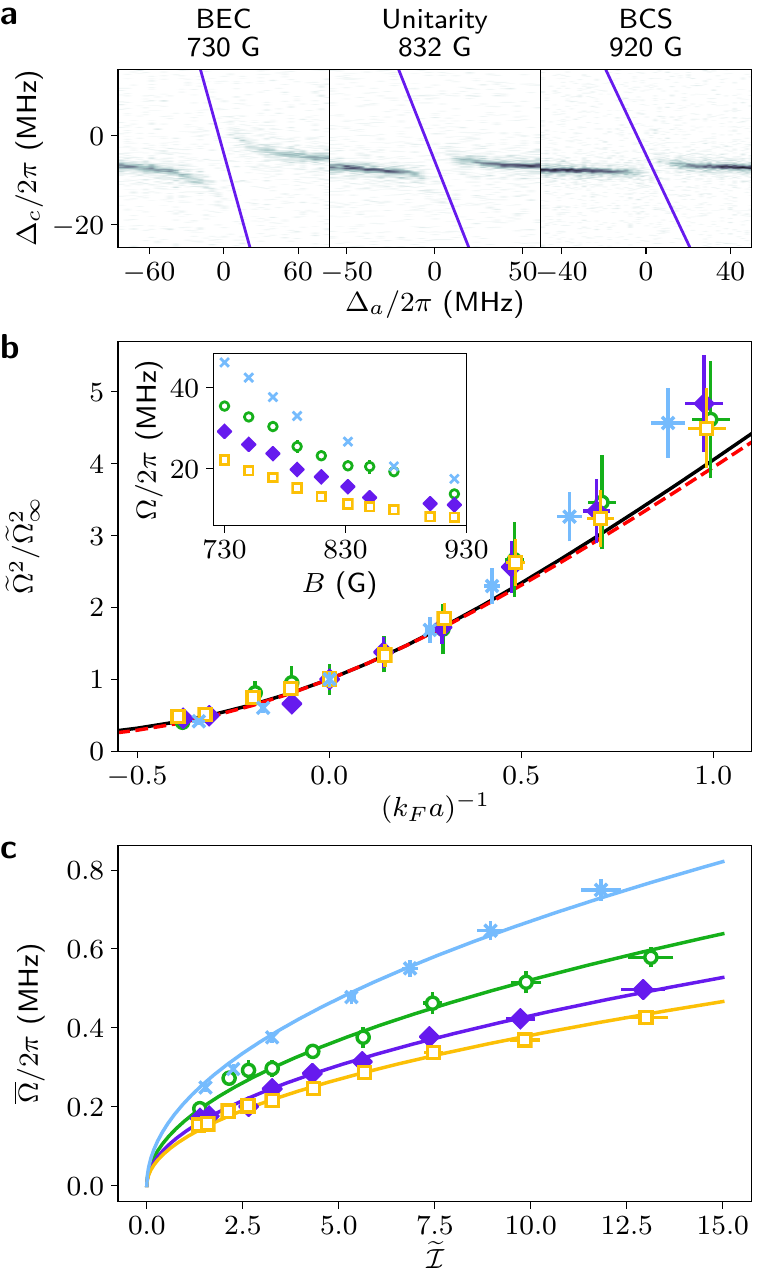}
	\caption{\textbf{Interaction dependent photon--pair coupling.} \textbf{a}~Transmission spectra of the PA resonance 2 in the BEC, unitarity and BCS regimes. $\Delta_a/2\pi$ is shifted by $-3.27$, $-3.20$ and $-3.14\,\si{GHz}$ for the BEC, unitarity and BCS, respectively, for visibility. The solid lines indicate the fitted location of the PA line. The color scale is identical to figure \ref{Fig2}. \textbf{b}~Magnetic field dependence of the collective Rabi frequencies $\Omega$ (inset) and the evolution of $\widetilde{\Omega}^2$ (see text), normalized by its value at unitarity $\widetilde{\Omega}_\infty^2$ for PA1 (green open circles), 2 (purple filled diamonds), 3 (orange open squares) and $1 ^3 \Sigma_g^+,v=81$ (light blue crosses) as a function of $1/k_Fa$. The black solid and red dashed lines represent the trap averaged contact calculated by Gaussian pair fluctuations and quantum Monte-Carlo, respectively, normalized by the value at unitarity. \textbf{c}~Scaling of $\overline{\Omega}$ (see text) as a function of the trap averaged contact $\tilde{\mathcal{I}}$ inferred from Gaussian pair fluctuations. The solid lines are square root fit to the data.
	}
	\label{Fig3}
\end{figure}

\begin{figure*}[t]
	\includegraphics{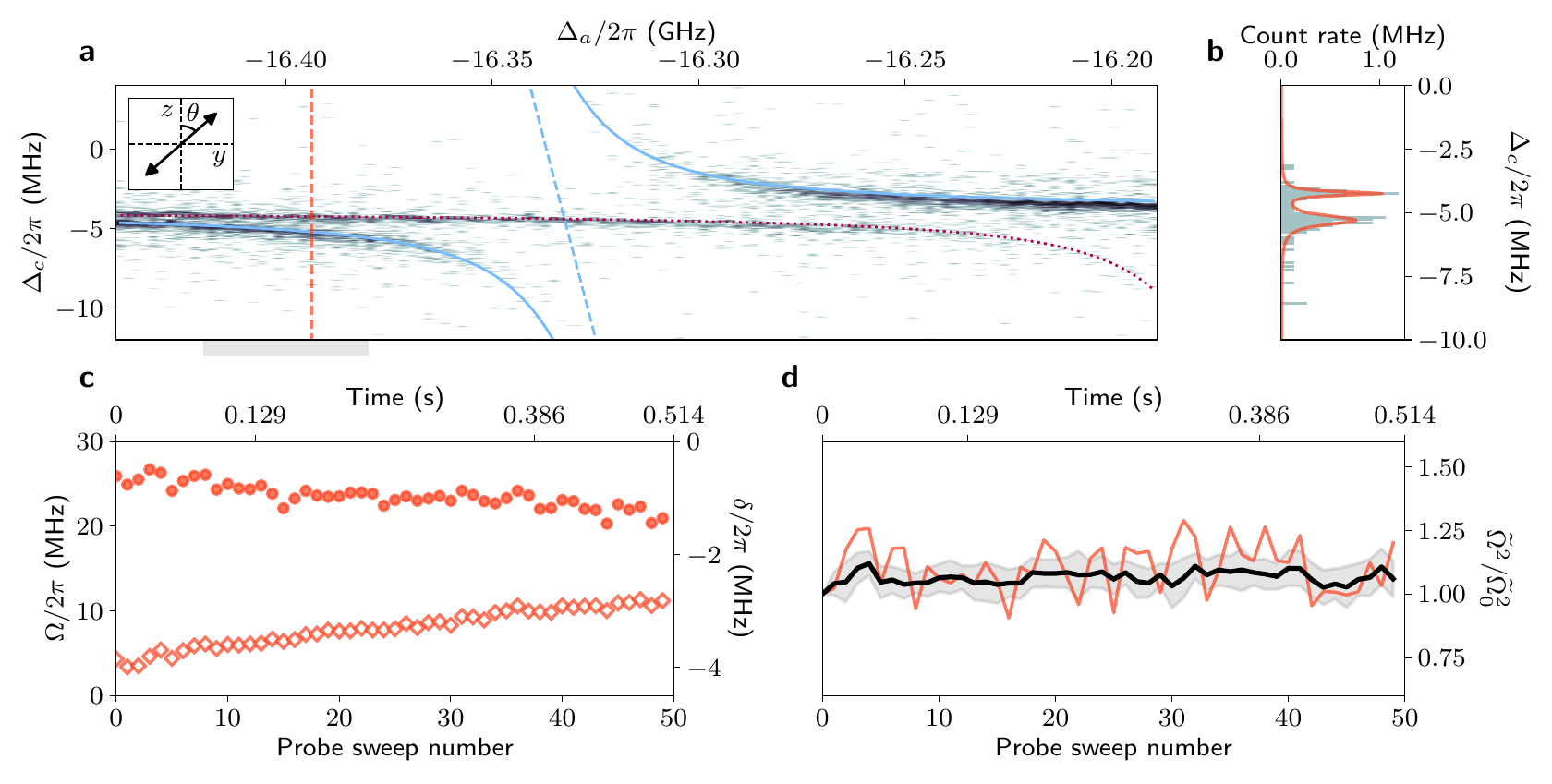}
	\caption{\textbf{Single-shot, repeated measurement of pair correlations.} \textbf{a}~Transmission spectrum close to the $1 ^3 \Sigma_g^+,v=81$ state, for a balanced Fermi gas composed of $8.0(2) \times 10^5$ atoms at $832\,\si{G}$ with a probe polarization tilted by $\theta=26\degree$ (inset). The two resonances around $\Delta_a/2\pi=-16.333$ and $-16.178\,\si{MHz}$ correspond to the $\pi$- and $\sigma^+$-polarization components, respectively. The dashed blue line indicates the position of the resonance coupled to $\pi$-polarized photons. The blue solid and the red dotted lines show the polariton positions of the $\pi$- and $\sigma^+$-polarized photons, respectively, extrapolated from a single measurement. The color scale is identical to figure \ref{Fig2}.
	\textbf{b}~Raw photon counts at $\Delta_a/2\pi=-16.394\,$\si{GHz} (orange dashed line on panel a). The solid line is a double-Lorentzian fit to the histogram. Two transmission peaks located at $-5.2$ and $-4.3 \,$\si{\mega\hertz} correspond to $\pi$ and $\sigma$ polarized photons respectively.
	\textbf{c}~Extracted Rabi frequency (filled circles) and atomic dispersive shift (open diamonds) from $50$ consecutive measurements on a single cloud at $\Delta_a/2\pi=-16.394\,$\si{GHz}.
	\textbf{d}~ Evolution of $\widetilde{\Omega}^2$ renormalized by its initial value $\widetilde{\Omega}_{0}^2$ evaluated from successive measurements for a single cloud (orange line) and averaged over $20$ different clouds (black solid line) for detunings ranging from $\Delta_a/2\pi=-16.380$ to $\Delta_a/2\pi=-16.420 \,\si{GHz}$ (shaded area on panel a). The grey area represents the standard deviation.
	}
	\label{Fig4}
\end{figure*}

The correlation function appearing in equation~\eqref{omega_main} depends on the shape and range of the molecular orbitals. We estimate that $R_c = 245\,a_0$, $210\,a_0$ and $180\,a_0$ with the Bohr radius $a_0$ for PA1, 2 and 3, respectively, such that $f(r)$ has a range much shorter than the inter-particle distance, the de Broglie wavelength and the photon wavelength (see Methods). In this regime, the correlation function has a universal dependence on the interaction strength, described by Tan's contact, the thermodynamic, many-body parameter canonically conjugated to interaction strength~\cite{Tan:2008aa,Zwerger:2012aa}.

We test the universality by measuring the evolution of the Rabi frequency throughout the BEC--BCS crossover. Figure~\ref{Fig3}a presents three typical spectra for PA2, at $1/k_Fa = 0.98$, $0.0$ and $-0.38$ ($B=730$, $832$ and $920$\,G), where $k_F$ and $a$ are the Fermi wave-vector and the $s$-wave scattering length, respectively. The avoided crossing  pattern characteristic of the strong coupling regime persists for all values of interaction strengths. The location of the PA resonance varies weakly and linearly with $B$ (see Methods). The Rabi frequency is larger on the BEC side, as can be seen in the inset of figure~\ref{Fig3}b, and smoothly decreases by a factor of about two towards the BCS regime.
We repeated this measurement for the two other PA resonances, as well as for the resonance coupling to state $1 ^3 \Sigma_g^+,v=81$ located at $\Delta_a /2\pi = -25.998\,\si{GHz}$, all showing qualitatively similar behaviors.

This suggests that the coupling to PA transitions is only dependent on the nature of the pairs in the ground state but not on the details of the molecular transition.
A calculation using a simplified orbital model shows that the variations of Rabi frequency are equal to that of Tan's contact up to a small correction accounting for the finite value of $R_c/a$ (see Methods). Specifically, we expect $\Omega^2 \propto \widetilde{\mathcal{I}} N k_F |1-\frac{R_c}{a}|^{2}$, where $\widetilde{\mathcal{I}}$ is the dimensionless, trap-averaged contact. Figure~\ref{Fig3}b presents the evolution of $\widetilde{\Omega}^2=\frac{\Omega^2}{Nk_F}|1-\frac{R_c}{a}|^{-2}$, normalized by its value at unitarity $\widetilde{\Omega}_\infty$, for the different PA lines. We observe the data collapse on each other, a striking manifestation of the universal character of the interaction dependence. The solid and dashed lines on figure~\ref{Fig3}b present the interaction dependence of the trap-averaged contact at $T=0$ calculated ab-initio in~\cite{Hu:2011aa,Hoinka:2013aa}, without any fit parameter, showing very good agreement with the data. We attribute the slight deviation in the BEC regime to the limited accuracy of the molecular orbital model. This demonstrates the direct correspondence between the optical spectrum of the cavity and the many-body physics of the quantum gas. Remarkably, the variations of free energy with interaction strength quantified by the contact are of the order of the Fermi energy in the ground state, but give rise to variations of Rabi frequency by $20\,$\si{\mega\hertz} from the BEC to the BCS regime, a magnification of the many-body effects by about $500$ in energy. 

The contact directly measures the number of pairs at short distance in a given volume of the gas. Equation~\eqref{omega_main} thus shows that the Rabi frequency scales with the square root of the number of pairs at a distance $R_c$, reminiscent of the scaling with the number of emitters in the Tavis--Cummings model. Having established the connection between the variations of $\Omega$ with interactions and the contact, we now consider the contact as known from theory and study the Rabi frequencies $\overline{\Omega}=\Omega/\sqrt{Nk_FL}$, where $L$ is the estimated width of the molecular wavefunction lobe around $R_c$ (see Methods), for four different PA transitions as a function of the contact, as shown in figure~\ref{Fig3}c. The dependence is well fitted by a square root function highlighting the coherent coupling of Fermion pairs with photons. Using a simplified model of the molecular orbital to connect the Rabi frequency with the number of molecules at distance $\sim R_c$ (see Methods), the fit provides an order-of-magnitude estimate for the single-photon--single-pair Rabi frequency $\Omega_0$. We obtain $\Omega_0/2\pi=826(12)$, $683(7)$, $604(6)\,\si{kHz}$ and $1.06(1)\,\si{MHz}$ for PA1, 2, 3 and $1 ^3 \Sigma_g^+,v=81$, respectively, comparable to the single-atom--single-photon Rabi frequency of $2\pi\times780\,$\si{\kilo\hertz} in the system. While pairs in the unitary gas are inherently a many-body effect, and cannot be isolated individually, we nevertheless conclude that our cavity has a cooperativity for single pairs and single photons approaching one. This suggests that all the quantum optics protocols designed in the context of single atom--photon interactions could be directly generalized to Fermion pairs in strongly interacting gases.

In contrast with other existing methods \cite{Kuhnle:2010aa,Sagi:2012aa,Chang:2016aa,Laurent:2017aa,Carcy:2019aa,Mukherjee:2019aa}, the combination of cavity QED with pairs through PA offers an avenue to measure the short range pair correlations dispersively with a single interrogation, allowing for time-resolved, repeated measurements on a single atomic sample.
To this end, we adapt our transmission spectroscopy technique to independently extract from a single probe sweep both the overall dispersive shift originating from the coupling to single atoms, and the extra contribution due to the coupling to a PA transition (see Methods). We rotate the probe polarization by $26\degree$ with respect to the magnetic field direction, such that the probe acquires a finite $\sigma^+$-polarization component. We probe the system close to transition to the state  $1 ^3 \Sigma_g^+,v=81$, the large detuning with respect to the atomic transitions mitigating spurious heating effects. There, the $\pi$ and $\sigma$ polarized photons couple to different excited molecular states at $\Delta_a/2\pi=-16.333$ and $-16.178\,\si{GHz}$ respectively, where $\Delta_a = 0$ is taken at the D1 $\pi$ transition. The spectrum taken in this configuration is shown in figure~\ref{Fig4}a for a unitary gas comprising $4.0(2) \times 10^5$ atoms per spin state, where each vertical line corresponds to a single interrogation on one cloud.

An example of raw photon counts obtained on a single interrogation is shown in figure~\ref{Fig4}b at $\Delta_a/2\pi=-16.378 \,\si{GHz}$. We fit the positions of the two resonances corresponding to the $\pi$ and $\sigma^+$ components using a double Lorentzian model. A typical spectrum corresponds to $88$ detection events which translate on average to $16$ intracavity photons on the transmission resonance, and of the order of one spontaneous emission event over the entire cloud. The positions of the resonances at a given $\Delta_a/2\pi$ allows us to retrieve both the atomic dispersive shift $\delta$ and the collective Rabi frequency of the molecular transition $\Omega$. The pair-polariton spectrum extrapolated from this single probe sweep is presented on figure~\ref{Fig4}a showing excellent agreement with a full data set covering all detunings.

Such a measurement sequence is then repeated $50$ consecutive times on the same cloud, separated by $10 \,$\si{\milli\second}, longer that the longest trap period, allowing for equilibration of the cloud between each interrogation. Figure~\ref{Fig4}c shows $\delta$ and $\Omega$, extracted from each measurement. The retrieved values of $\Omega$ reflect the time-evolution of two-body correlations for one single atomic cloud. We observe a decrease of $\Omega /2\pi$ from $26$ to $20 \,$\si{\mega\hertz} over the $50$ consecutive measurements, and a decrease of $\delta$ by $27\%$ over the consecutive measurements, the latter reflecting atom losses. About half of these losses can be attributed to the finite lifetime of the cloud  \cite{Roux:2020ab}. The measurement-induced atom losses are thus $1.9(1) \times 10^3$ atoms per sweep, $0.2 \%$ of the initial atom number. 

To confirm that the many-body physics is preserved in spite of the repeated measurements, we evaluate $\widetilde{\Omega}^2= \Omega^2 / Nk_F$ at each point in time. The result is shown in figure ~\ref{Fig4}d, normalized by its initial value, showing no discernible decay, where the noise originates predominantly from photon shot noise. An average over $20$ different traces measured on different clouds for several values of $\Delta_a$ is shown in figure~\ref{Fig4}d. This quantity is directly proportional to the trap averaged contact $\widetilde{\mathcal{I}}$, and remains constant demonstrating the weak heating originating from the measurements. Independent temperature measurements (see Methods) indicate an initial temperature of $0.09(1) T_F $ and a final one of $0.15(1)T_F $ after the $50$ probe sweeps, for which we would expect a decrease of the trap-averaged contact by about $7 \%$ \cite{Kuhnle:2011aa}.

The ability to observe in time pair-correlations while preserving the many-body physics is an ideal starting point for future theoretical and experimental investigations of quantum noise and back-action mechanisms for correlation measurements in many-body systems. It opens the fascinating perspective of combining quantum-limited sensing with strongly correlated matter. We expect the quantum noise spectra of such measurements to carry fundamental information on high order correlations, similar to the case of simple atom number measurements~\cite{Uchino:2018ab}. 

In addition, the ability of the cavity field to couple directly to the pair correlation function in the dispersive regime suggests the possibility to engineer pair--pair interactions mediated by cavity photon exchanges \cite{Ritsch:2013aa}, opening an uncharted territory to quantum simulation. Beyond these fundamental questions, the weakly destructive and time-resolved character of the cavity-assisted measurement will be of immediate, practical interest in the study of correlations after quenches, such as spin diffusion \cite{Enss:2019aa}, repulsively interacting Fermi gases where pairing competes with ferromagnetism \cite{Amico:2018aa}, or during slow transport processes~\cite{Krinner:2017aa,Zeiher:2020aa}, complementing other high-efficiency methods \cite{Eisele:2020aa}. Last, our work adds the exquisite control over photons of a high finesse cavity to the existing cold molecules toolbox~\cite{Bohn:2017aa},  opening the way to dissipation engineering of cold chemistry~\cite{Perez-Rios:2017aa,Kampschulte:2018aa,Wellnitz:2020aa}.

\bibliography{PA}
\bibliographystyle{naturemag}

\section{Acknowledgements}
We thank Tobias Donner for discussions and a careful reading of the manuscript, Randy Hulet, Paul Julienne and Jeremy Hutson for discussions, Chris Vale, Hui Hu and Joaquin Drut for providing the contact data and Timo Zwettler for experimental assistance. We acknowledge funding from the European Research Council (ERC) under the European Union's Horizon 2020 research and innovation programme (grant agreement No 714309), the Swiss National Science Foundation (grant No 184654), the Sandoz Family Foundation-Monique de Meuron program for Academic Promotion and EPFL.

\section{Authors contributions}
HK, KR and VH performed the experiments and processed the data, HK, KR and JPB wrote the paper, JPB planned and supervised the project. HK and KR contributed equally. 

\section{Methods}
\renewcommand{\figurename}{Extended Data Fig.}
\setcounter{figure}{0} 

\subsection{Experimental procedure}
We produce a quantum degenerate, strongly interacting Fermi gas of $^6$Li following the method described in~\cite{Roux:2020ab}. Atoms captured in a magneto-optical trap are first transferred into a standing wave optical dipole trap at $1064\,\si{\nm}$ created by the TEM$_{01}$ mode of the cavity. Prior to evaporative cooling the magnetic field is ramped to $832\,\si{G}$, the location of the broad Feshbach resonance between the two lowest hyperfine states $\ket{1}$ and $\ket{2}$. Evaporative cooling is first performed in the cavity before transfer to a crossed optical dipole trap, formed by two running wave beams at $1064\,\si{nm}$ with a waist of $33\,\si{\micro\m}$ intersecting at an angle of $22^\circ$ at the center of the cavity. This yields a spin-balanced unitary Fermi gas containing $4.6\times10^5$ atoms with a temperature of $T=0.07T_F$ with the Fermi temperature $T_F$.

The cloud is recompressed in $50\,\si{ms}$ followed by $50\,\si{ms}$ hold time in a trap with frequencies of $2\pi\times(273,619,624)\,\si{Hz}$, with the weakest confinement along the cavity direction, before the cavity interrogation. This avoids unwanted trap losses and ensures maximal coupling of the cloud with the TEM$_{00}$ mode of the cavity. The recompression is accompanied by a magnetic field ramp from $832\,\si{G}$, where the whole evaporation is performed, to various values in order to change the inter-atomic interaction.

We send probe beam pulses at $671\,\si{nm}$ mode-matched to the TEM$_{00}$ mode of the cavity as depicted in figure~\ref{Fig1}a. The probe beam is linearly polarized and we set it either along or tilted from the magnetic field orientation depending on the measurements we perform (see the main text). Each pulse is $500$-$\si{\micro s}$ long and frequency-swept over $40\,\si{MHz}$, which controls $\Delta_c/2\pi$. Transmission from the cavity is collected on a single photon counter. At the end of an experimental sequence a reference probe pulse measures the cavity resonance without atoms to compensate for drifts.

\subsection{Fit to the spectrum}

The transmission of light through a cavity in the presence of atoms takes the form~\cite{Tanji-Suzuki:2011ac}
\begin{equation*}
T = \frac{1}{\bigl(1+H\frac{\Omega^2}{\kappa\gamma}\mathcal{L}_a(\delta)\bigr)^2+\bigl(\frac{2\Delta_c}{\kappa}+H\frac{\Omega^2}{\kappa\gamma}\mathcal{L}_d(\delta)\bigr)^2},
\end{equation*}
with the probe--atom detuning $\delta$, the probe--cavity detuning $\Delta_c$, the cavity field decay rate $\kappa$, the atomic decay rate $\gamma$, the collective Rabi frequency $\Omega$ and the collective coupling parameter $H=1/2$ for a uniformly distributed atomic ensemble. The absorption and dispersion parts of the scattering $\mathcal{L}_a(\delta)$ and $\mathcal{L}_d(\delta)$ are given by
\begin{equation*}
\mathcal{L}_a(\delta)=\frac{1}{1+4(\delta/\gamma)^2}~~\mathrm{and}~~\mathcal{L}_d(\delta)=\frac{-2\delta/\gamma}{1+4(\delta/\gamma)^2}.
\end{equation*}
With the definition of the main text (figure~\ref{Fig2}), we have $\delta = \Delta_c+\Delta_a$.

The observed avoided crossings arise from the coupling between the atom--photon dressed state and the molecular bound states. For the fit to these avoided crossings $\kappa$ and $\gamma$ should represent the decay rates of the atomic polariton and the molecular state, respectively. We further introduce a linear slope coefficient $\alpha$ to account for the detuning dependence of the dispersive shift due to the atomic transition over the spectrum window. This leads to a fit function with an amplitude parameter $q$:
\begin{widetext}
\begin{equation}
T_{fit}=\frac{q}{\bigl(1+\frac{\Omega^2}{2\kappa\gamma}\mathcal{L}_a(\Delta_c+\Delta_a)\bigr)^2+\bigl(\frac{2(\Delta_c-\alpha\Delta_a)}{\kappa}+\frac{\Omega^2}{2\kappa\gamma}\mathcal{L}_d(\Delta_c+\Delta_a)\bigr)^2}.
\label{Tfit}
\end{equation}
\end{widetext}

\begin{figure}[t]
	\includegraphics{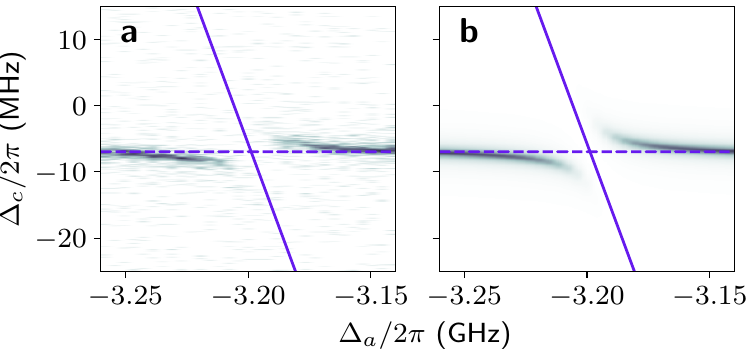}
		\caption{\textbf{Fit to the spectrum.} \textbf{a}~Spectrum of PA2 at $832\,\si{G}$ averaged over three realizations shown in figure~\ref{Fig2}. \textbf{b}~Spectrum reconstructed by equation~\eqref{Tfit} using the fit results. The solid and dashed lines indicate the fitted positions of the PA resonance and the dispersively shift cavity resonance. The color scale is identical to that of the main text.}
	\label{fit}
\end{figure}

To extract error bars for the parameters fitted over the maps of figure~\ref{Fig2}c--e and \ref{Fig3}a, we repeat each measurement on three different samples. We first construct a spectrum by randomly choosing one scan among the three realizations for each $\Delta_a$ and fit it with equation~\eqref{Tfit}. We then repeat this procedure $100$ times for each set of parameters. The mean and the standard deviation of the fit results represent the value and error bars shown in figure~\ref{Fig3} and mentioned in the text. Extended data figure~\ref{fit} shows an example of the fits for PA2 at $832\,\si{G}$.

\begin{figure}[t]
	\includegraphics{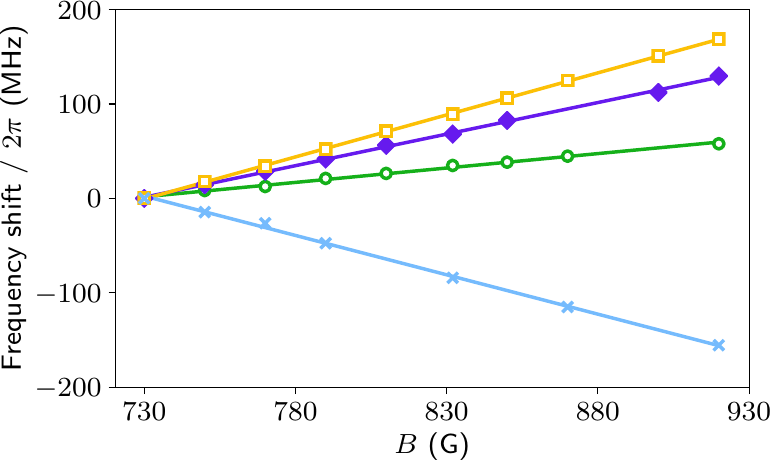}
	\caption{\textbf{Magnetic field dependence of the binding energies.} Positions of the photoassociation resonances PA1, 2, 3 and $1^3\Sigma_g^+$, (green open circles, purple filled diamonds, orange open squares and light blue crosses, respectively) with respect to the atomic D2 $\pi$ (for PA1--3) and D1 $\pi$ (for $1^3\Sigma_g^+,v=81$) transitions at each magnetic field. The value at $730$ G is subtracted for clarity. Linear fits presented by the solid lines yield $0.31$, $0.67$, $0.89$ and $-0.83\,\si{MHz/G}$ for PA1, 2, 3 and $1^3\Sigma_g^+$, respectively.}
	\label{papos}
\end{figure}

Offsets in $\Delta_c$ and $\Delta_a$ (omitted in equation~\eqref{Tfit}) are used to determine the PA position and the overall dispersive shift $\delta_{at}$ due to the coupling with single atoms. The PA positions linearly depend on magnetic fields by $0.31$, $0.67$, $0.89$ and $-0.83\,\si{MHz/G}$ for PA1, 2, 3 and $1^3\Sigma_g^+$, respectively, as shown in extended data figure~\ref{papos}.

Far detuned from atomic transitions, the dispersive shift is given by $\delta_{at}=\tilde{N}g_0^2/\Delta_a$ with $\tilde{N}$ the atom number coupled to the cavity field and the single-atom--single-photon coupling strength $g_0$ ~\cite{Tanji-Suzuki:2011ac}. As most PA transitions used in our experiment are located between the D1 and D2 $\pi$ transitions, which are $10\,\si{GHz}$ apart, we take contributions from the both transitions into account to infer the total atom number $N$ as
\begin{equation*}
N=\frac{2\delta_{at}/\zeta}{g_{D1\pi}^2/\Delta_{D1\pi}+g_{D2\pi}^2/\Delta_{D2\pi}}.
\end{equation*}
Here the coupling strengths of the two transitions are $g_{D1\pi}/2\pi=276\,\si{kHz}$ and $g_{D2\pi}/2\pi=390\,\si{kHz}$ and $\Delta_{D1\pi}$ and $\Delta_{D2\pi}$ are the detunings from each transition at the location of the resonance. The factor two accounts for the standing wave structure of the cavity field and $\zeta\sim0.95$ is a factor for mode overlap between the cloud and the cavity field in the transverse direction.

\subsection{Theoretical description}
To connect the vacuum Rabi splitting observed on the photo-association line with the pair correlation function, we start with a general single-mode light--matter Hamiltonian in the rotating wave approximation
\begin{equation*}
\hat{H} = \delta \hat{a}^\dagger \hat{a} + \frac{i\Omega_0}{2} \left( \hat{F}^\dagger \hat{a} - \hat{F} \hat{a}^\dagger \right),
\end{equation*}
where $\delta$ is the detuning between the cavity and atomic transition, $\hat{a}$ annihilates a cavity photon and $ \hat{F}$ is an operator acting on the atomic and molecular Hilbert space describing the interaction process. In the Heisenberg picture, the second derivative of the cavity field operator reads
\begin{equation*}
\left( \frac{\partial^2}{\partial t^2} + \delta^2 + \frac{\Omega_0^2}{4}\left[\hat{F},\hat{F}^\dagger\right]  \right) \hat{a}= i \delta \frac{\Omega_0}{2}\hat{F}.
\end{equation*}
For zero detuning, taking expectation values and using a mean-field decoupling between light and matter, the normal mode splitting can be directly read out as
\begin{equation}
{\Omega}^2 =  \Omega_0^2\left\langle\left[\hat{F},\hat{F}^\dagger\right] \right\rangle.
\label{omega}
\end{equation}

For a photo-association process, a minimal model of the light--matter coupling interaction part can be written as 
\begin{equation*}
\hat{F}^\dagger = \int  d{\bf R}  d{\bf r} g({\bf R}) f({\bf r}) \hat{\psi}_m^\dagger({\bf R}) \hat{\psi}_1({\bf R} - \frac{{\bf r}}{2}) \hat{\psi}_2({\bf R} + \frac{{\bf r}}{2}),
\end{equation*}
where $\hat{\psi}_\sigma({\bf r})$ is a field operator annihilating a Fermion in state $\sigma$ at position ${\bf r}$, and $\hat{\psi}_m({\bf R})$ annihilates a molecule with a center of mass at position ${\bf R}$. $g(R)$ is the mode function of the cavity, and $f(r)$ describes the relative motion of atoms in the target molecular state. We evaluate the expectation value in equation~\eqref{omega} under the low saturation hypothesis $\left\langle  \hat{\psi}_m^\dagger({\bf R})  \hat{\psi}_m({\bf R})  \right\rangle \sim 0$, to obtain equation~\eqref{omega_main} in the main text.

\subsection{Molecular orbitals}
The molecular states we target are bound in the $2S +2P$ asymptotic potential. We work with $\pi$-polarized light, at strong magnetic fields and the binding energies of PA1, 2 and 3 are much smaller than the fine structure splitting of lithium and of the order of the Zeeman shift. Therefore, for these states we use a minimal model of the molecular potential originating from the D2 $\pi$-transition only, with transition dipole moments computed from the Breit--Rabi formula. 

The validity of this model can be tested using the Leroy--Bernstein formula connecting the van-der-Waals coefficient $C_3$ to the location of the highly excited vibrational bound states. For the three photo-association lines PA1, 2 and 3 used in the main text, a fit leaving as single free parameter the location of the continuum describes the binding energies better than $5\%$. Note however that for most of the transitions we observed, in particular closer to the D1 line or with other polarization we could not reproduce the spectrum with such a simple model. For the $1 ^3 \Sigma_g^+,v=81$ orbital we directly use the documented $C_3$ coefficient to calculate the Condon radius \cite{Abraham:1995aa}.

\subsection{Universality}
The connection between the pair correlations in equation~\eqref{omega_main} and Tan's contact is standard. Following~\cite{Zwerger:2012aa}, we write the pair correlation function as 
\begin{equation*}
\left\langle \hat{\psi}_1^\dagger(r_1) \hat{\psi}_2^\dagger(r_2)   \hat{\psi}_2(r_2) \hat{\psi}_1(r_1) \right \rangle = \sum_i n_i \phi^{(i)*}(r_1,r_2)\phi^{(i)}(r_1,r_2),
\end{equation*}
with $n_i$ real positive coefficients and $\phi^{(i)}(r_1,r_2)$ a set of short-range, normalized two-body orbitals that obey the Bethe--Peierls boundary conditions 
\begin{equation*}
\phi^{(i)}(r_1,r_2) \xrightarrow[|r_1 - r_2| \rightarrow 0]{} A^{(i)}\left( \frac{1}{|r_1 - r_2|} - \frac{1}{a} \right),
\end{equation*}
 where $A^{(i)}$ is a normalization coefficient and $a$ is the scattering length. Therefore, we get 
\begin{multline*}
\left\langle \hat{\psi}_2^\dagger({\bf R} + \frac{{\bf r}'}{2}) \hat{\psi}_1^\dagger({\bf R} - \frac{{\bf r}'}{2})   \hat{\psi}_1({\bf R} - \frac{{\bf r}}{2}) \hat{\psi}_2({\bf R} + \frac{{\bf r}}{2}) \right \rangle \\ = \sum_i n_i({\bf R}) |A^{(i)}({\bf R})|^2 \left( \frac{1}{|r|} - \frac{1}{a} \right)\left( \frac{1}{|r'|} - \frac{1}{a} \right),
 \end{multline*}
in the relevant range of relative distances. We introduce the contact as 
\begin{equation*}
\mathcal{C}({\bf R}) = 16 \pi^2 \sum_i n_i({\bf R}) |A^{(i)}({\bf R})|^2,
\end{equation*}
and thus express the commutator as
\begin{multline*}
\left\langle\left[ \hat{F}, \hat{F}^\dagger \right] \right\rangle \\= \int  d{\bf R} |g({\bf R})|^2 \mathcal{C}({\bf R},a) \left| \int_0^\infty dr \chi(r) \left( 1 - \frac{r}{a} \right) \right|^2,
\end{multline*}
where we have explicitly written the scattering-length dependence of the contact. We have also specialized to $s$-wave molecular states and introduced $\chi(r) = rf(r)$. The molecular state is independent of interaction strength in the ground state. 

In the strongly interacting regime where the Condon radius $R_c$ is much smaller than the scattering length, the dependence of the Rabi frequency on scattering length is fully universal, captured by the contact. The term $L=\left| \int_0^\infty dr \chi(r) \right|^2$ is a length representing the width of the outer lobe of the molecular orbital close to the Condon point. The commutator is then simply the number of pairs of Fermions $N_p$, with inter-particle distance within an interval of length $L$ centered around $R_c$. The overall coupling to the field is modulated by the mode function, such that on average only half the pairs contribute, and the Rabi splitting reads
\begin{equation}
\Omega =  \Omega_0 \sqrt{\frac{N_p}{2}}.
\label{eq:omegascaling}
\end{equation}
This is the expected behavior of the set $N_p/2$ of identical emitters coherently coupled to one mode of the field. Our experiment thus represents the photo-association counterpart of the celebrated Tavis--Cummings model. 

In order to account for the finite value of $R_c$ in the theory--experiment comparison, we follow \cite{Cote:1995aa} and model the target molecular orbital as a square box of width $L$ centered around $R_c$, such that the integral can be evaluated explicitly
\begin{equation*}
\left\langle\left[ \hat{F}, \hat{F}^\dagger \right] \right\rangle = \frac{L}{4\pi}\left| 1 - \frac{R_c}{a} \right|^2 \int  d{\bf R} |g({\bf R})|^2 \mathcal{C}({\bf R}).
\end{equation*}
Therefore, we expect the evolution of the Rabi splitting with scattering length, compared with the value measured at unitarity to obey 
\begin{equation*}
\frac{\Omega^2(a)}{\Omega^2(\infty)} = \frac{\int  d{\bf R} |g({\bf R})|^2 \mathcal{C}({\bf R},a)}{\int  d{\bf R} |g({\bf R})|^2 \mathcal{C}({\bf R},\infty)} \cdot \left| 1 - \frac{R_c}{a} \right|^2.
\end{equation*}

%The theory line 
The data points in figure~\ref{Fig3} includes the finite Condon radius correction, estimated from the $C_3$ coefficient and the binding energies for the four different transitions. For data in the far BEC regime, the term accounting for the finite size of the molecule contributes by about $15\textrm{--}20\%$ to the scaling.

\subsection{Number of short range molecules} 
In order to connect the scaling with the contact observed in the experiment to the single-atom--single-pair coupling, we estimate the number of pairs in the gas addressed by the PA. To this end, we evaluate the length $L$ introduced in the above theory. This can be done by supposing that the molecular potential close to $R_c$ is approximately linear with the position $r$ as
\begin{equation*}
V(r) = -E_b \left( 1 - \frac{3r}{R_c} \right),
\end{equation*}
where we have expressed the $C_3$ coefficient in terms of the binding energy $E_b$. Therefore the molecular orbital is approximately an Airy function, with a width $L = \left(\frac{\hbar^2 R_c}{3 m E_b}\right)^{1/3}$, where $m$ is the atomic mass and $\hbar$ is the Planck constant divided by $2\pi$. We obtain thus $L = 28.9a_0$, $23.6a_0$, $19.4a_0$ and $12.6a_0$ for PA1, 2, 3 and $1^3\Sigma_g^+,v=81$, respectively. 

By virtue of Tan's relations, for a distance $s$ much smaller than other many-body length scales, the number of pairs $dN_p$ with a volume $d^3R$ centered around point ${\bf R}$ is 
\begin{equation*}
dN_p({\bf R},s) = \frac{\mathcal{C}({\bf R}) s}{4 \pi} d^3R
\end{equation*}
Integrating over the whole cloud we get the total number of pairs at distance $s$ in terms of the integrated contact $\mathcal{I}$:
\begin{equation*}
N_p(s) = \frac{\mathcal{I} s}{4 \pi} = \mathcal{\tilde{I}}\frac{N k_F s}{4 \pi}
\end{equation*}
with $\mathcal{I} = \mathcal{\tilde{I}} N k_F$, and $N$ the total atom number. Note that the factor of $2$ due to the averaging over the $\cos^2$ mode function has already been incorporated in the scaling at the level of equation~\eqref{eq:omegascaling}.

\subsection{Single shot estimate of the Rabi frequency}

We model the position of the dressed states close to a PA transition by

\begin{equation*}
E_{\pm} = -\frac{\hbar(1-\alpha)(\Delta_a-\Delta_\mathrm{PA})}{2}\pm \frac{\hbar \sqrt{\Omega^2 +(\Delta_{a}-\Delta_\mathrm{PA})^2}}{2} + \hbar\delta_{at},
\end{equation*}
with $\Omega$ the Rabi frequency, $\alpha$ the linear slope originating from the dressing of the cavity with the atomic transition, $\Delta_\mathrm{PA}$ the position of the PA transition with respect to the atomic transition and $\delta_{at}$ the atom-induced dispersive shift. We found that both dressed states emerging from each polarization component fit this model with the same Rabi frequency.

Accounting for the presence of one resonance for each polarization, we determine $\alpha$, $\delta_{at}$ and $\Omega$ from the measurement of the position of the two transmission resonances.

%A single trace of $\Omega$ gives the evolution of the two body correlation function over time for a single cloud. Taking several traces to calculate $\langle \Omega^2 \rangle/\langle Nk_F\rangle$ over different sample at different value of $\Delta / 2\pi$ allows to observe the time evolution of Tan's Contact while the cloud is interrogated $50$ times.

\end{document}